\documentclass[aps,pra,showpacs,twocolumn,amsmath,amssymb,foo
tinbib]{revtex4}

\usepackage[english]{babel}
\usepackage{times}
\usepackage{latexsym}

\usepackage{graphicx} 
\usepackage{epsf} 
\usepackage{subfigure} 
\usepackage{amsmath} 
\usepackage{amssymb} 
\usepackage{amsfonts}
\usepackage{bm}
\usepackage{natbib}
\usepackage{epstopdf}

\def\be{\begin{equation}}
\def\ee{\end{equation}}
\def\bea{\begin{eqnarray}}          
\def\eea{\end{eqnarray}}
\def\bi{\begin{itemize}}
\def\ei{\end{itemize}}
\def\bin{\begin{enumerate}}
\def\ein{\end{enumerate}}

\begin{document}

\title{Characterizing the Hofstadter butterfly's outline with Chern numbers}


\author{
N. Goldman
}
\affiliation{Center for Nonlinear Phenomena and Complex Systems - Universit\'e Libre de Bruxelles (U.L.B.), Code Postal 231, Campus Plaine, B-1050 Brussels, Belgium}

\date{\today}

\begin{abstract}

In this work, we report original properties inherent to independent particles subjected to a magnetic field by emphasizing the existence of regular structures in the energy spectrum's outline. We show that this fractal curve, the well-known Hofstadter butterfly's outline, is associated to a specific sequence of Chern numbers that correspond to the quantized transverse conductivity. Indeed the topological invariant that characterizes the fundamental energy band depicts successive stairways as the magnetic flux varies. Moreover each stairway is shown to be labeled by another Chern number which measures the charge transported under displacement of the periodic potential. We put forward the universal character of these properties by comparing the results obtained for the square and the honeycomb geometries.
\end{abstract}


\maketitle

\section{Introduction}

The energy spectrum associated to charged particles moving in a two-dimensional lattice and subjected to a high magnetic field has inspired numerous works, since the seminal papers of Hofstadter \cite{Hofstadter} and Wannier \cite{Wannier}. When the magnetic flux penetrating the lattice is a rational number, namely $\Phi=p/q$ where $p,q$ are integers, the spectrum splits into $q$ subbands. The representation of the spectrum as a function of the flux shows a recursive structure with clear self-similarities \cite{MacDonald1983}, and adopts the shape of an intriguing insect: the so-called Hofstadter butterfly \cite{Hofstadter}. The infinitely many gaps which compose this surprising figure are known to follow a simple rule: each gap is labeled by two integers $(t_r,s_r)$ which satisfy a Diophantine equation \cite{Wannier,Kohmoto1992}. In a fundamental work, Thouless \emph{et al.} have emphasized that the Hofstadter butterfly  plays a key role in the quantum Hall effect theory: when the Fermi energy of the system lies in a gap, the transverse conductivity of the system is quantized and is given by $\sigma _{xy}= (e^2/h) \, t_r$, where $e$ is the particle's charge, $h$ is Planck's constant and $t_r$ satisfies the aforementioned Diophantine equation \cite{Thouless}.\\
In this context, the Green-Kubo expression for the transverse conductivity has an elegant topological interpretation: when the Fermi energy of the system lies in a gap, the transverse conductivity is given by a sum of Chern numbers. The latter are associated with the filled bands that are situated below the gap \cite{Kohmoto1985}. From a mathematical point of view, the Chern number is an integral invariant characterizing the topology of a fibre bundle which is defined on each energy band \cite{Simon1983,Kohmoto1985}. Moreover, these topological invariants are known to be the \emph{only} quantized quantities which can be associated with the energy bands \cite{Avron1983}. The Diophantine equation, which describes any 2D electronic systems subjected to a magnetic field \cite{Kohmoto1992}, is highly connected to topology since both integers $(t_r,s_r)$ can be interpreted as Chern numbers \cite{Kunz1986,Halperin1989}. As pointed out by MacDonald and Kunz, the Chern number $s_r$ measures the charge transported when the periodic potential is adiabatically displaced \cite{MacDonald1983,Kunz1986}. \\

Recent advances in cold atoms physics \cite{Bloch,ouradv} allow for the experimental exploration of the Hofstadter butterfly using atoms trapped in optical lattices \cite{Jaksch}. Different arrangements, which indeed mimic the presence of an artificial ``magnetic" field in the dynamics of neutral particles \cite{Ho2001,Polini2005,Jaksch,Demler,Jaksch2}, are nowadays studied in laboratories \cite{porto}. With such setups one expect to explore new features in the fields of vortex physics \cite{ouradv, Holland1, Goldman2,dan1} and quantum Hall systems \cite{Demler,Palmer,Goldman1, Holland2,Lew,Palmer2}. In particular, we have suggested that an integer quantum Hall-like effect for neutral fermionic particles should be observed in optical lattices \cite{Goldman1}. \\
Recently, we have also shown that fermionic atoms trapped in 2D optical lattices and subjected to an artificial ``magnetic" field should undergo a Mott metal-insulator transition and that the phase boundary depicts the Hofstadter butterfly's \emph{outline} \cite{Goldman3}. Although the Mott-insulator phase transition occurs in the system when the interaction between the particles is taken into account, the phase boundary only depends on the underlying single-particle physics. A similar result has been obtained by Oktel \emph{et al.} \cite{oktel1,oktel2} and by Goldbaum \emph{et al.} \cite{dan} in the context of the bosonic superfluid-insulator transition. \\
Motivated by the important role played by the Hofstadter butterfly's outline in this theoretical framework, and by the recent experimental advances in the field of ultracold atoms, we investigate intrinsic properties associated to this fractal and intriguing curve. We point out that the Hofstadter butterfly's outline is also known to represent the phase boundary for the normal-superconducting phase transition in superconducting networks \cite{lopez,rammal,alex}. \\

In this paper, we characterize the Hofstadter butterfly's outline by studying  the Chern numbers associated to the fundamental energy band of the system as a function of the magnetic flux. Physically, we evaluate the system's transverse conductivity  when the Fermi energy lies in the first gap of the energy spectrum. Under this condition, we show that the measure of the transverse conductivity exhibits a particular sequence of plateaus as a function of the magnetic flux. We study such structures and emphasize their universal character by comparing the results obtained for the square and for the honeycomb geometries.

\section{The square lattice}

\begin{center} 
\begin{figure}
{\scalebox{0.32}{\includegraphics{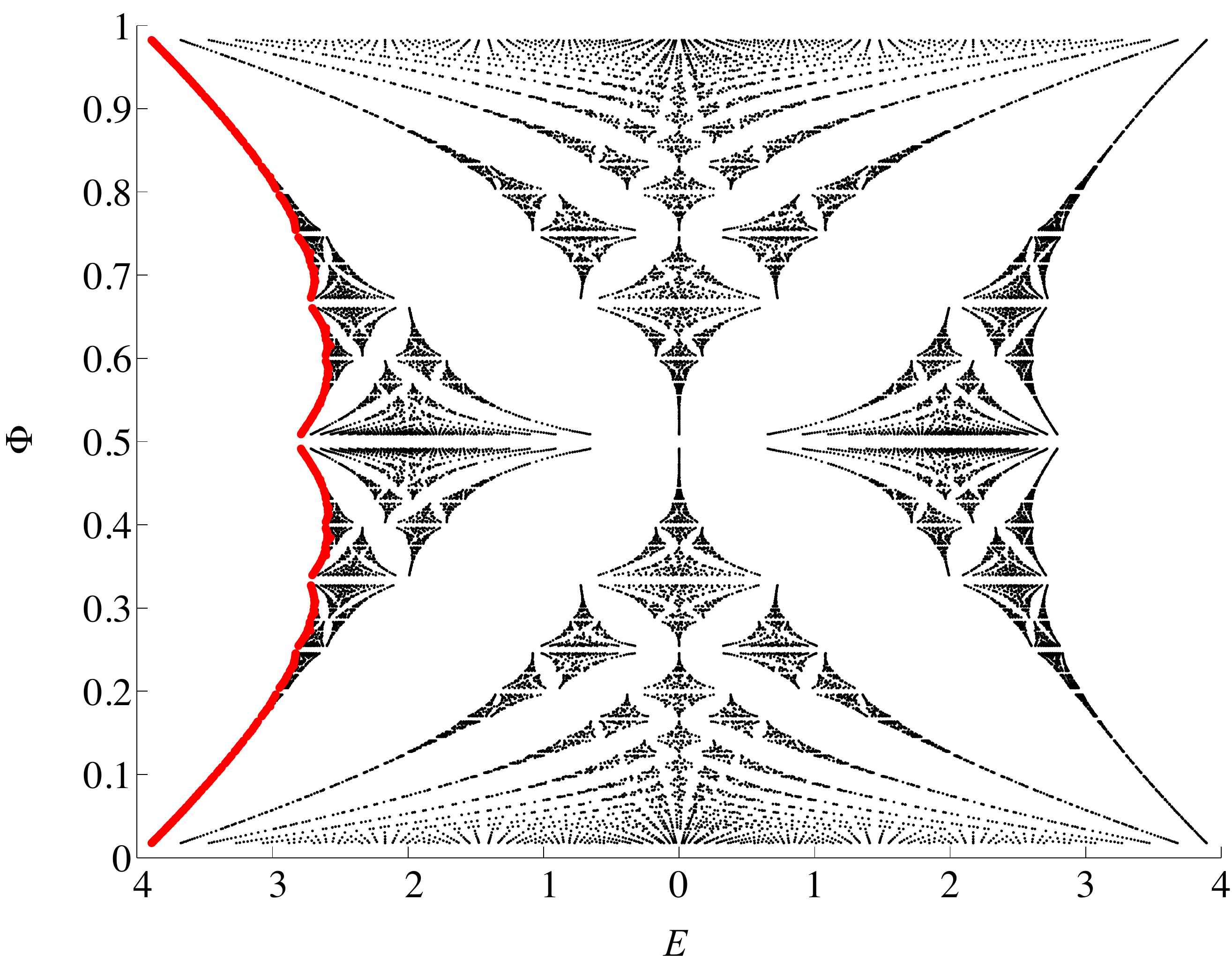}}} 
\caption{\label{square_but} (Color online) Hofstadter butterfly $\Phi= \Phi(E)$: single-particle spectrum for a 2D square lattice subjected to a magnetic field. The butterfly's outline, which corresponds to the fundamental energy band $E_1(k_x,k_y)$, is colored in red. The energy is expressed in units of the hopping parameter $t$. }
\end{figure} 
\end{center}

We first consider the case of a fermionic gas trapped in a 2D optical \emph{square} lattice and subjected to an artificial ``magnetic" field \cite{ouradv,Jaksch}. We assume that the optical potential created by the lasers is sufficiently strong in order to apply a tight-binding approximation and we investigate this system in the non-interacting limit, which can be reached at low densities. The magnetic field $\boldsymbol{B}=B \, \boldsymbol{1_z}$, which is characterized by the number of effective magnetic flux quanta per unit cell $\Phi$, is supposed to be induced by rotation \cite{Holland1,Holland2}, or by combining laser-assisted tunneling and lattice acceleration methods \cite{Jaksch,Demler,porto} or by immersion of the lattice into a rotating Bose-Einstein condensate \cite{Jaksch2}. For a rotating system, this parameter is given by $\Phi=2 M a^2 \Omega /h$, where $\Omega$ is the angular velocity, $M$ is the particle's mass, $a$ is the lattice constant and $h=2 \pi \hbar$ is Planck's constant \cite{Holland2}. We can treat this problem by considering the Landau gauge, $\boldsymbol{A}=(0, B x,0)$, for which the many-body Hamiltonian reads $\mathcal{H}=  t \sum_{m,n} \mathcal{H}_{m,n}$, with
\begin{align} 
\mathcal{H}_{m,n}=  a^{\dagger}_{m,n} a_{m-1,n}  + e^{i 2 \pi \Phi m} a^{\dagger}_{m,n} a_{m,n-1} + {\rm h.c.} ,
\label{ham}
\end{align}
where $a_{m,n}$ ($a^{\dagger}_{m,n}$) is the fermionic annihilation (creation) operator on the lattice site $(m,n)$ and $t$ is the nearest-neighbor tunneling amplitude. In this work, the effective magnetic flux $\Phi$ is supposed to vary between $[0,1]$, which is achieved for $\Omega \approx 1$ kHz. In the following, we set the lattice constant $a$ and the particle's mass $M$ to unity and work in units where $\hbar=1$ except otherwise stated. For rational fluxes, $\Phi=p/q$ where $p$ and $q$ are integers, the single-particle Schr\"odinger equation yields the well-known Harper equation \cite{Hofstadter}
\begin{equation}
e ^{i k _x} \, \psi_{m+1} + e ^{-i k _x} \, \psi_{m-1}  + 2 \cos (2 \pi \Phi m - k _y) \, \psi_m = \frac{E}{t}  \, \psi_m ,
\label{harp}
\end{equation}
where $\psi_{m}$ is a $q$-periodic wave function, $\boldsymbol{k}$ is the wave vector and $E$ is the single particle energy. The wave vector belongs to the magnetic Brillouin zone, a two-torus defined as $k_x \in [0, \frac{2 \pi}{q}]$ and  $k_y \in [0,2 \pi]$. The energy spectrum associated to Eq. \eqref{harp} has a band structure, composed of $q$ subbands, which has been extensively studied in the literature \cite{Pets,teo,komo,MacDonald1983,Hofstadter,Wannier}: the representation of the energy as a function of the flux leads to the Hofstadter butterfly (Fig.\ref{square_but}). This striking fractal figure, which illustrates the infinitely many gaps of the spectrum, exhibits a recursive structure and is described through a simple rule \cite{MacDonald1983}: for $\Phi =p/q$, the $r^{th}$ gap of the spectrum is labeled by two integers $(t_r , s_r)$, which satisfy a Diophantine equation \cite{Wannier, Kohmoto1992}
\begin{equation}
r = p t_r + q s_r .
\label{diop}
\end{equation}
In the square lattice case, the condition $\vert t_r \vert \le q/2$ determines the solution unambiguously \cite{Kohmoto1992}.

In this work, we focus our attention on the outline of the butterfly (red curve in Fig.\ref{square_but}), which is known to play a key role in the field of quantum phase transitions \cite{Goldman3,oktel1,oktel2,dan,lopez,rammal,alex}. Our aim is to investigate the structures characterizing this non-trivial curve, which is formed by the fundamental band $E_1 (k_x, k_y)$, where $H \vert \psi _1\rangle=E_1\vert \psi_1 \rangle$, in the range $\Phi =p/q  \in [0,1]$. As $q$ increases, the butterfly's outline doesn't seem to smoothen because of its fractal nature \cite{Hofstadter}.  In order to maintain the system along this irregular curve, we suppose that the Fermi energy lies inside the first gap of the single particle spectrum for all $\Phi \in [0,1]$. Under this assumption, we are able to compute the analogue of the transverse conductivity for neutral currents \cite{Goldman1},  which is associated to the energy band $E_1(k_x, k_y)$ for a given value of the flux $\Phi$.  In this context, the transverse conductivity can be computed with Kubo's formula and is expressed as \cite{Thouless,komo}

\be
\sigma _{xy}=\frac{1}{h} \frac{1}{2 \pi i} \int_{\mathbb{T}^2} \langle \frac{\partial \psi_1}{ \partial k_x} \vert \frac{\partial \psi_1}{ \partial k_y} \rangle- \langle \frac{\partial \psi_1}{ \partial k_y} \vert \frac{\partial\psi_1}{ \partial k_x} \rangle ,
\label{cond3}
\ee
where the fundamental state $\vert \psi_1 \rangle$ alone contributes.

The quantization of this quantity follows from the topological interpretation of Eq. \eqref{cond3}: the transverse conductivity is related to the topologically invariant Chern number $C_S$ \cite{remark}, an integer defined as
\begin{align}
C_S&= \frac{i}{2 \pi } \int_{\mathbb{T}^2} \mathcal{F}  \notag \\
&=\frac{i}{2 \pi} \int_{\mathbb{T}^2} \langle \frac{\partial \psi_1}{ \partial k_x} \vert \frac{\partial \psi_1}{ \partial k_y} \rangle- \langle \frac{\partial \psi_1}{ \partial k_y} \vert \frac{\partial\psi_1}{ \partial k_x} \rangle ,
\end{align}
where $\mathcal{F}$ is the so-called Berry's curvature associated to the band $E_1(k_x, k_y)$. According to Refs. \cite{Thouless,komo}, one finds that $C_s= - t_1$, and therefore this invariant integer satisfies the Diophantine equation Eq.\eqref{diop} with $r=1$, 

\be
C_S=\Phi^{-1} s_1 - \frac{1}{p},
\label{dio1}
\ee
with the condition $\vert C_S \vert \le q/2$. \\

\begin{center} 
\begin{figure}
{\scalebox{0.47}{\includegraphics{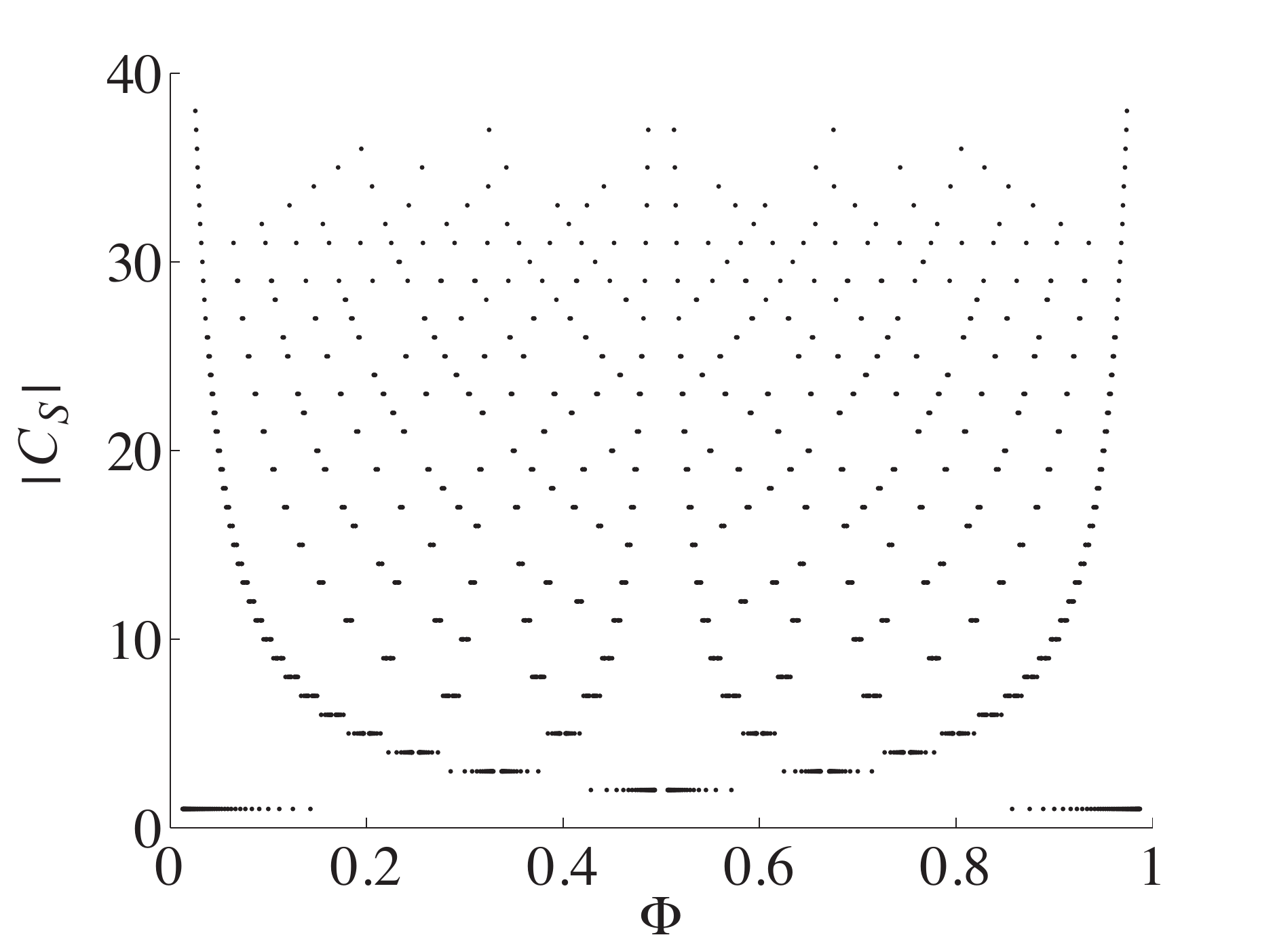}}} 
\caption{\label{two} Chern numbers as a function of the magnetic flux, $\vert C_S \vert= \vert C_S \vert (\Phi)$,  for $\Phi= \frac{p}{q}$ with $q<77$. }
\end{figure} 
\end{center}

One can solve Eq.\eqref{dio1} for all $\Phi=p/q \in [0,1]$, in order to obtain the many Chern numbers associated to the Hofstadter butterfly's outline. Technically one fixes a high value for the denominator $q$ and computes the Chern number $C_S$ for $p=1, 2,...,q$, such that $p$ and $q$ are mutually primes. \\
The illustration of these integers as a function of the effective magnetic flux is quite surprising. All along the irregular outline $E_1=E_1(\Phi)$, the Chern numbers computed for the various $\Phi$ follow a very regular law: the representation of the Chern numbers as a function of the flux $C_S= C_S (\Phi)$ depicts plateau sequences, adopting the shape of successive stairways. In Fig.\ref{two}, we show this structure in a compact way, by plotting $\vert C_S \vert$ as a function of the flux $\Phi$. We note that this figure is symmetric with respect to $\Phi=0.5$.

\begin{center} 
\begin{figure}
{\scalebox{0.31}{\includegraphics{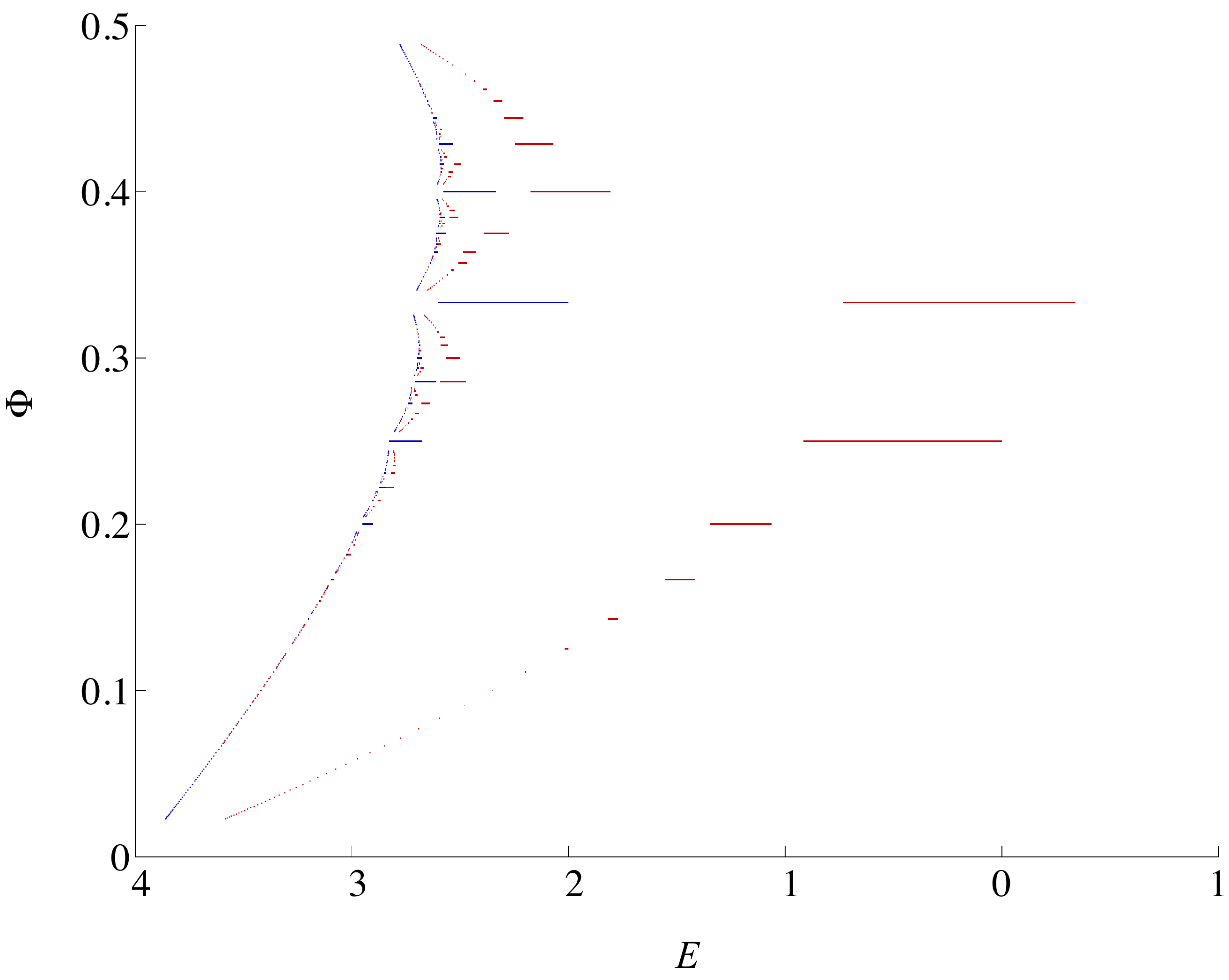}}} 
\caption{\label{edges}  Two lowest energy bands, $E_1$ (blue lines) and $E_2$ (red lines), as a function of the flux $\Phi$. The energy is expressed in units of the hopping parameter $t$.}
\end{figure} 
\end{center}

The numerical result illustrated in Fig. \ref{two} is already interesting since it underlines the complexity of the Hofstadter butterfly's outline. A topological argument stipulates that if the fundamental band $E_1 (k_x,k_y)$ is well separated from the first excited state $E_2 (k_x,k_y)$, for all $\Phi \in [0,1]$, then the Chern number $C_S$ should remain constant \cite{Avron1983}. However, this is not the case since we have shown that this topological number takes many different values as the flux varies. For $q$ fixed and $p=1,...q$, we obtain that the number of different values $N_{C}$ increases linearly with respect to the denominator $q$. We eventually note that the Chern number's magnitude $\vert C_S \vert $ takes the value of all the natural numbers up to $N_C$, namely $\vert C_S \vert=1,2,...,N_C$, and that $N_C \approx q/2$.  Since $q$ may take an arbitrarily high value, our calculations show that the first gap closes infinitely many times as the flux is varied. These numerous gap closings are illustarted in Fig. \ref{edges}, where the two lowest energy bands are plotted as a function of the flux  $\Phi$.\\

\begin{center} 
\begin{figure}
{\scalebox{0.29}{\includegraphics{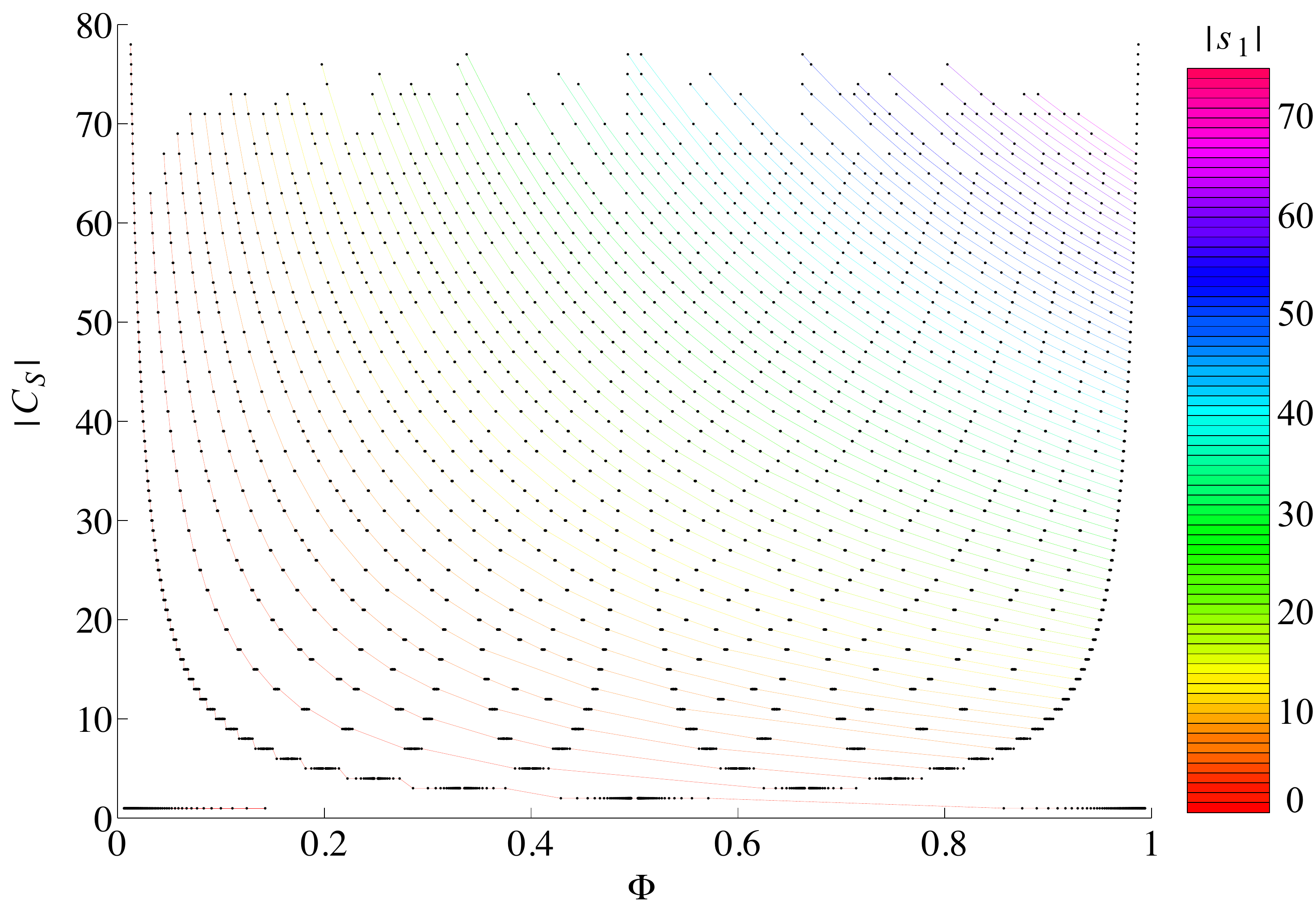}}} 
\caption{\label{three} (Color online) Chern numbers (black dots) as a function of the magnetic flux, $\vert C_S \vert= \vert C_S \vert (\Phi)$, for $ \Phi= \frac{p}{q}$ with $q<157$. The dots are connected by colored lines according to the topological invariant $\vert s_1 \vert$ to which they are associated through the Diophantine equation Eq.\eqref{dio1}. Note that the Chern numbers are single-valued.}
\end{figure} 
\end{center}

Furthermore, we find that a strong structure is hidden behind the intriguing Fig.\ref{two}. We show that the other Chern number which satisfies the Diophantine equation with $t_1= - C_S$, 
\be
s_1=\frac{1}{q} + \Phi C_S,
\label{dio2}
\ee
classifies the many points $(\Phi, \vert C_S \vert)$ plotted in Fig.\ref{two}  in a rigorous manner.  We illustrate this result in Fig.\ref{three}, in which we connect all the points $(\Phi, \vert C_S \vert)$ that are associated to the same number $\vert s_1 \vert$ with a colored line. As $\vert s_1 \vert$ increases, the color changes progressively from red to purple. It is very clear from this colored figure that all the plateaus belong to a specific stairway, labeled by $\vert s_1 \vert$, since no crossing between the lines is observed. Moreover, we point out that the successive stairways correspond to increasing values of $\vert s_1 \vert$. We note that the integer $t_1$ is related to the quantized transverse conductivity, whereas the number $s_1$ actually measures the charge transported when the periodic potential is adiabatically displaced \cite{MacDonald1983,Kunz1986},\\

In summary, the first gap of the spectrum closes infinitely many times as the magnetic flux is varied. A signature of these gap closings is given by the Chern number $C_S$, which is associated to the first energy band, and which takes infinitely many different values as a function of the flux. Surprisingly, these many values give rise to a structure consisting of successive stairways which are characterized by the number $s_1$.

\section{The Honeycomb lattice}

\begin{center} 
\begin{figure}
{\scalebox{0.15}{\includegraphics{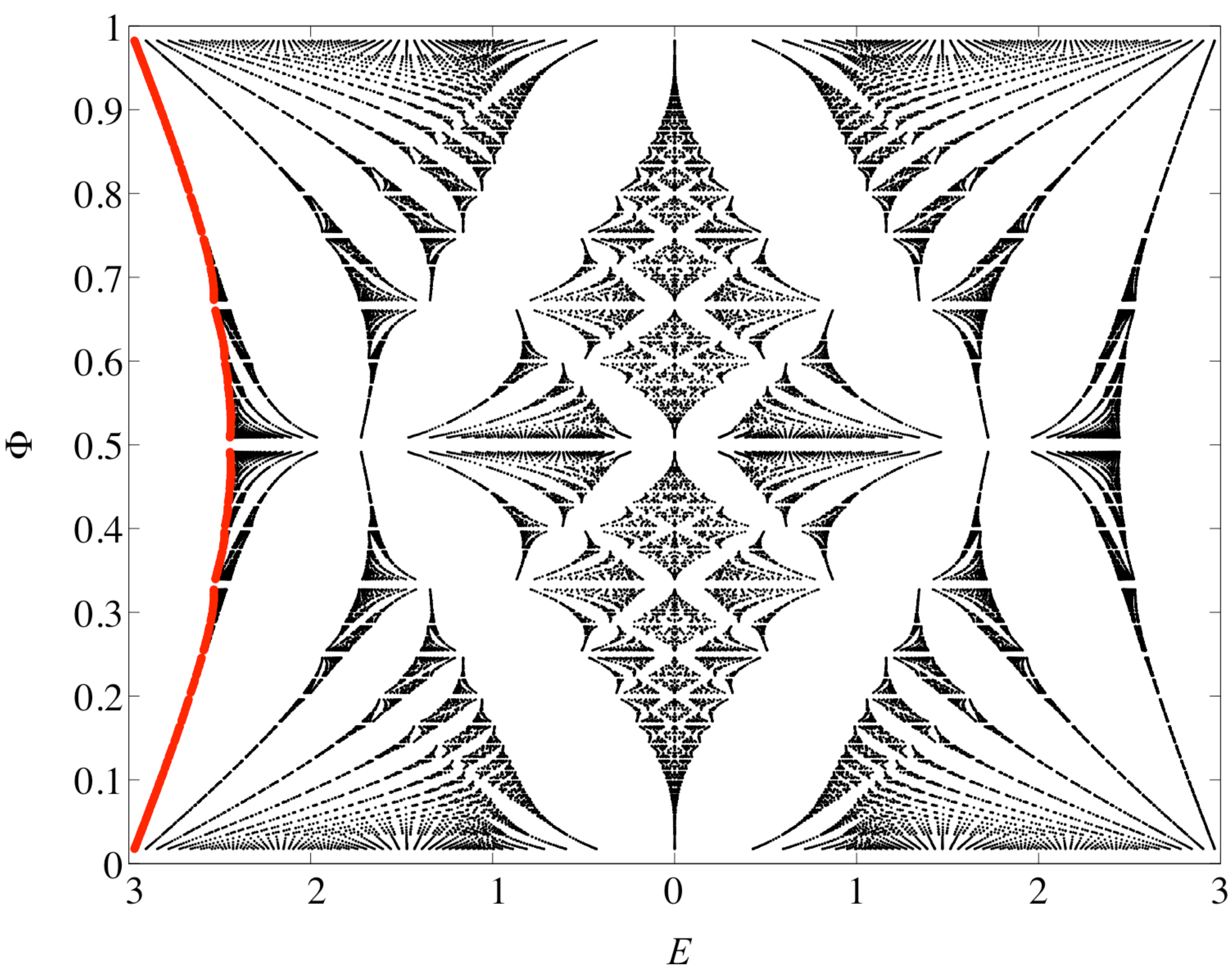}}} 
\caption{\label{honey_but} (Color online) Butterfly spectrum $\Phi= \Phi(E)$ for the 2D honeycomb lattice subjected to a magnetic field. The butterfly's outline is colored in red. The energy is expressed in units of the hopping parameter $t$.}
\end{figure} 
\end{center}

In this section, we consider the case of a 2D \emph{honeycomb} lattice subjected to a magnetic field. The honeycomb has a bipartite structure and it is common to define two fermion operators $a^{\dagger}_A (\boldsymbol{r})$ and $a^{\dagger}_B (\boldsymbol{r})$, where $\boldsymbol{r}=m \boldsymbol{e}_1 + n \boldsymbol{e}_2$ (see for exemple Ref.\cite{graphene}). The unit vectors are chosen as $\boldsymbol{e}_1=(3/2, \sqrt{3}/2)$ and $\boldsymbol{e}_2=(0, \sqrt{3})$.  In the tight-binding approximation, the many-body Hamiltonian reads $\mathcal{H}=  t \sum_r \mathcal{H}(\boldsymbol{r})$, where
\begin{align} 
&\mathcal{H}(\boldsymbol{r})=  a^{\dagger}_A (\boldsymbol{r}) a_B(\boldsymbol{r})  + e^{i 2 \pi \Phi m} a^{\dagger}_A(\boldsymbol{r}) a_B(\boldsymbol{r}-\boldsymbol{e}_2) \notag \\
&+ a_A^{\dagger}(\boldsymbol{r}+\boldsymbol{e}_1) a_B(\boldsymbol{r})  +{\rm h.c.} ,
\label{hamhoney}
\end{align}
and $\Phi=p/q$ is the effective magnetic flux quanta per unit cell. The single-particle Schr\"odinger equation associated to Eq.\eqref{hamhoney} yields

\begin{align} 
& \biggl (1+e^{+ i 2 \pi \Phi m-i k_y} \biggr ) \psi_A(m) +e^{i k_x}  \psi_A(m+1)=\frac{E}{t} \psi_B (m) , \notag \\
& \biggl (1+e^{-i 2 \pi \Phi m+i k_y} \biggr ) \psi_B(m) +e^{-i k_x}  \psi_B(m-1)=\frac{E}{t} \psi_A (m) , 
\label{hamhoney2}
\end{align}
where $\psi_A(m)$ and $\psi_B(m)$ are $q$-periodic wave functions, $\boldsymbol{k}$ is the wave vector and $E$ is the single-particle energy. The wave vector belongs to the magnetic Brillouin zone
 defined as $k_x \in [0, \frac{2 \pi}{q}]$ and  $k_y \in [0,2 \pi]$. The spectrum is depicted in Fig.\ref{honey_but} as a function of the effective magnetic flux $\Phi$, and illustrates a modified version of the Hofstadter butterfly. The band structure associated to Eq.\eqref{hamhoney2} has been extensively studied in order to investigate the very rich physics of graphene \cite{graphene,Sharapov2005,Park2008}. The Chern numbers associated to the energy bands have been computed numerically by Hatsugai \emph{et al.} and give rise to the anomalous quantum Hall effect: around $E=0$, the transverse conductivity evolves by steps according to $\sigma_{xy}=\pm (2 N+1) e^2/h$ where $N$ is an integer \cite{graphene}.

\begin{center} 
\begin{figure}
{\scalebox{0.28}{\includegraphics{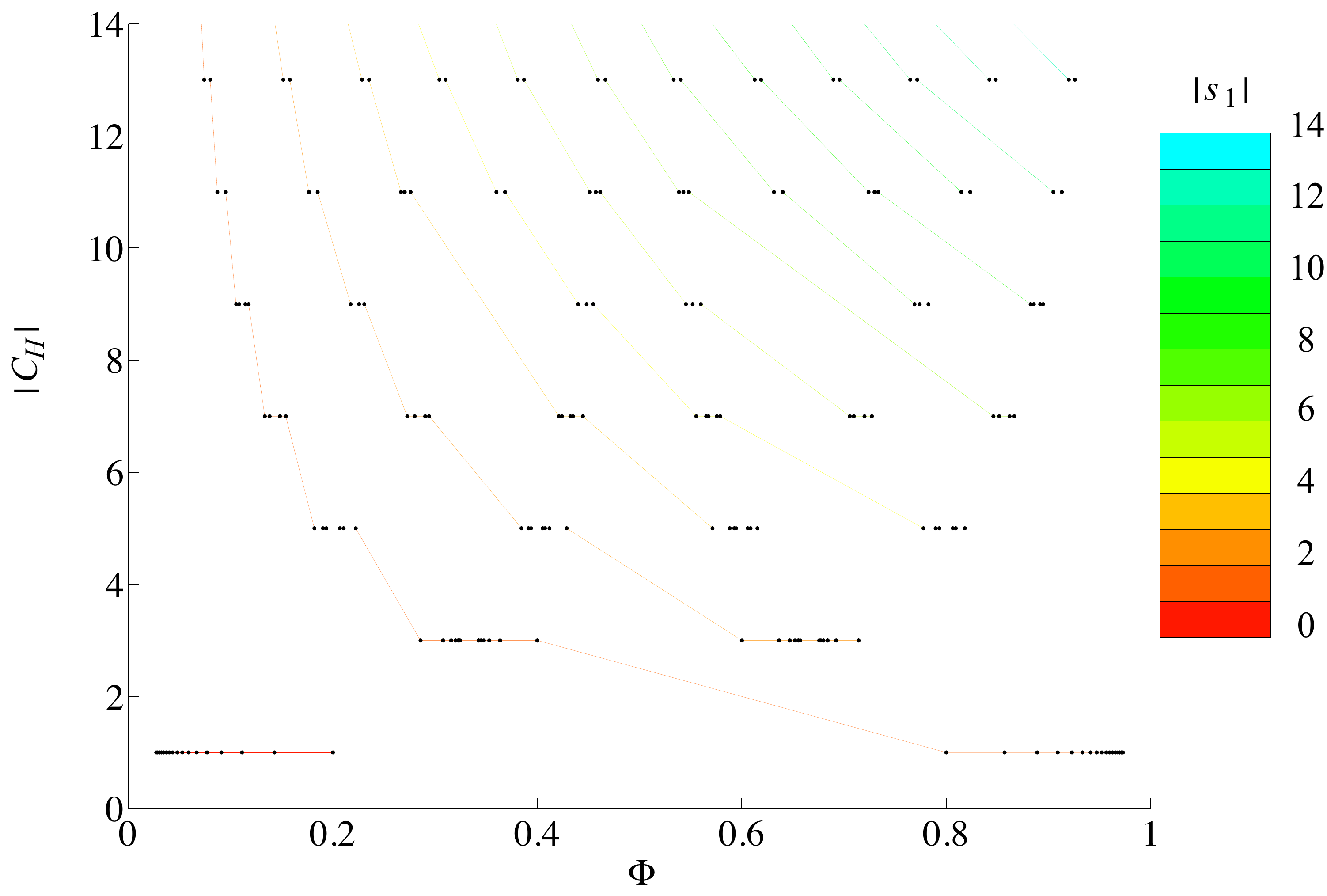}}} 
\caption{\label{honeychern} (Color online) Chern numbers (black dots) as a function of the magnetic flux, $\vert C_H \vert= \vert C_H \vert (\Phi)$, for $ \Phi= \frac{p}{q}$ with $q<37$. The dots are connected by colored lines according to the topological invariant $\vert s_1 \vert$ to which they are associated through the Diophantine equation Eq.\eqref{diop}. Note that the Chern numbers are single-valued.}
\end{figure} 
\end{center}

The outline of the honeycomb butterfly (red curve in Fig.\ref{honey_but}) is highly irregular and differs from the square lattice case in regards to its general shape. In order to characterize this other irregular curve, and compare it to the Hofstadter butterfly's outline, one has to evaluate the Chern numbers $C_H$ associated to the fundamental band $E_1(k_x,k_y)$ for the honeycomb lattice case. Unfortunately, contrary to the square lattice case, no Diophantine equation is known to describe the entire honeycomb butterfly \cite{graphene}: the gaps are labeled by two integers $(t_r,s_r)$ which satisfy Eq. \eqref{diop}, but the condition $\vert t_r \vert \le q/2$ is not always satisfied. Thus the Chern numbers cannot be unambiguously determined on the basis of the Diophantine equation. In the lack of such an equation, one has to compute the Chern numbers  numerically. This can be easily achieved thanks to an efficient method developed by Fukui \emph{et al.} \cite{fukui}. \\
The numerical results are shown in Fig.\ref{honeychern}. A new sequence of plateaus is observed in the honeycomb case, for which the transverse conductivity evolves by steps according to $\sigma_{xy}=\pm (2 N+1) e^2/h$, where $N$ is an integer.  This is already surprising, because the fundamental band $E_1(k_x,k_y)$ corresponds to energies which are far from $E=0$, where the anomalous behavior of double steps is expected \cite{graphene}. From this observation, it seems that the main signature of the anomalous quantum Hall effect is already contained in the edge of the butterfly: if the Fermi energy of the system lies in the first gap while varying $\Phi$, one should observe this fascinating property proper to graphene. From a topological point of view, this particular sequence of Chern numbers suggests that the honeycomb butterfly's outline radically differs from the Hofstadter butterfly's outline.\\

 However, it is worth noticing that the general structure depicted in Fig.\ref{honeychern} is remarkably similar to the results shown in Fig.\ref{two} and Fig.\ref{three} in the context of the square lattice case. The universality arising from these figures indicates that the Diophantine equation also plays a key role in the honeycomb system. \\
Although the condition $\vert C_H \vert \le q/2$ is not fulfilled, we have verified that all the solutions $C_H(\Phi)$ satisfy the Diophantine equation Eq. \eqref{diop}, and that each stairway is again characterized by the other number $\vert s_1 \vert$.\\
 In order to emphasize the universality suggested by Fig.\ref{two} and Fig.\ref{honeychern}, we have connected with a colored line the solutions $C_H(\Phi)$ according to their associated number $\vert s_1 \vert$ (see Fig.\ref{honeychern}). As in the square lattice case, the successive stairways are indeed characterized by increasing values of the Chern number $\vert s_1 \vert$. 
 
\section{Conclusion}

In this work, we have put forward the existence of strong and regular structures associated to the highly irregular Hofstadter butterfly's outline. These structures, which arise in square and honeycomb lattices, have a universal character and are related to the underlying topology of the system's fundamental energy band. The topology of the system is designated through the Chern number which gives its value to the quantized transverse conductivity.  The correspondence between the results obtained for the square and the honeycomb lattices has been confirmed through numerical computations of the Chern numbers in the honeycomb case. We have verified that these solutions indeed satisfy the general Diophantine equation. It has been shown, for both geometries, that when the Fermi energy remains in the first gap, the transverse conductivity is highly irregular but evolves on stairways labeled by the other number that satisfies the Diophantine equation.  We believe that these structures might play a role in the Mott-insulator transitions observed in rotating atomic systems, where single-particle properties are known to be dominant. We eventually point out that the properties emphasized in this work are not restricted to the field of cold atoms physics and could also be found in superconducting networks or 2D electronic systems. \\

N. G. thanks  P. Gaspard, A. Astudillo Fernandez, P. de Buyl, S. Goldman, R. Matos Alves, J.-S. Mc Ewen, N. Tabti and V. Wens for their support. The author also thanks A. Kubasiak and M. Lewenstein for their valuable comments and encouragements. N. G. is financialy supported by the F.R.S.-FNRS Belgium.



\begin{thebibliography}{99}
\bibitem{remark} We use the notation $C_S$ [resp. $C_H$] in order to denote the Chern numbers computed in the square [resp. honeycomb] lattice case.
\end{thebibliography}
\end{document}